\documentclass[aps,pre,twocolumn,showpacs,preprintnumbers,amsmath,amssymb]{revtex4}
\usepackage{epsfig}
\usepackage{epstopdf}
\usepackage{lipsum}

\begin{document}
\title{Partial equivalence of statistical ensembles in a simple spin model \\with discontinuous phase transitions}
\author{Agata Fronczak, Piotr Fronczak and Grzegorz Siudem}
\affiliation{Faculty of Physics, Warsaw University of Technology,
Koszykowa 75, PL-00-662 Warsaw, Poland}
\date{\today}

\begin{abstract}
In this paper, we draw attention to the problem of phase transitions in systems with locally affine microcanonical entropy, in which partial equivalence of (microcanonical and canonical) ensembles is observed. We focus on a very simple spin model, that was shown to be an equilibrium statistical mechanics representation of the biased random walk. The model exhibits interesting discontinuous phase transitions that are simultaneously observed in the microcanonical, canonical, and grand canonical ensemble, although in each of these ensembles the transition occurs in a slightly different way. The differences are related to fluctuations accompanying the discontinuous change of the number of positive spins. In the microcanonical ensemble, there is no fluctuation at all. In the canonical ensemble, one observes power-law fluctuations, which are, however, size-dependent and disappear in the thermodynamic limit \cite{2016aPREFronczak}. Finally, in the grand canonical ensemble, the discontinuous transition is of mixed-order (hybrid) kind with diverging (critical-like) fluctuations \cite{2016bPREFronczak}. In general, this paper consists of many small results, which together make up an interesting example of phase transitions that are not covered by the known classifications of these phenomena.
\end{abstract}
\maketitle

\section{Introduction}

Classic ideal gas has the same properties regardless of whether it is studied in the formalism of the microcanonical, canonical or grand canonical ensemble \cite{bookGibbs,bookHuang,bookReihel}. Many other models of statistical physics also show such a correspondence. Because of that time, over time, the feature began to be considered as a kind of paradigm of statistical physics. However, the truth is that the so-called \textit{equivalence of ensembles} does not hold in general and the prominent counterexamples are nonadditive systems, which include systems with long range interactions \cite{2009PhysRepCampa, bookCampa}.

The first mentions about nonequivalent ensembles began to appear about fifty years ago (e.g.~\cite{1968LyndenBell, 1970Thirring, 1971Hertel}), and they concerned astrophysical self-gravitating systems. Over time, however, the problem was also noticed in other physical systems and models (to name a few examples, see \cite{1999LyndenBell, 1999PRETorcini, 2001PRLBarre, 2004PhysAEllis, 2005PRLMukamel, 2007PhysACampa, 2010PRLKastner, 2013JStatMechMori, 2017PREHovhannisyan, 2018PREBaldovin}). Some time ago, systematic research on these phenomena started, which, in addition to studying specific models, laid the foundations for the general theory of nonequivalent ensembles \cite{2000JStatPhysEllis, 2003PhDTouchette, 2004PhysATouchette, 2005JStatPhysBouchet, 2009PhysRepTouchette}. 

Among the references just mentioned, two contributions \cite{2000JStatPhysEllis, 2005JStatPhysBouchet} deserve special attention. The first one, Ref.~\cite{2000JStatPhysEllis}, entitled \textit{Large deviation principles and complete equivalence and nonequivalence results for pure and mixed ensembles} due to Ellis et al., provides the basis for a complete mathematical theory of the problem. The authors show there, that the issue of nonequivalent ensembles can be resolved by examining the concavity of entropy as a function of energy, $S(U)$. In short, in Ref.~\cite{2000JStatPhysEllis} (see also \cite{2003PhDTouchette, 2004PhysATouchette}), the authors show that microcanonical and canonical ensembles are \textit{equivalent} when $S(U)$ is strictly concave. \textit{Nonequivalence} is observed when $S(U)$ is convex. Finally, the \textit{partial equivalence} is referred to systems with locally affine $S(U)$.

In turn, Ref.~\cite{2005JStatPhysBouchet}, entitled \textit{Classification of phase transitions and ensemble inequivalence in systems with long range interactions}, by Bouchet and Barre, takes up an important issue of phase transitions in nonequivalent ensembles. By combining the singularity and concavity analysis of the entropy $S(U)$, authors present a kind of thermodynamic classification of the phase transitions in nonequivalent ensembles. 

To be concrete, in the case of equivalent ensembles, microcanonical macrostates with the fixed energy $U$ and microcanonical temperature given by $\beta_{_m}\!=\!\partial S/\partial U$ directly correspond to canonical macrostates with the fixed temperature and average energy satisfying: $\beta=\beta_{_m}$ and $\langle U\rangle=U$, respectively. The equivalence of ensembles holds whenever entropy is a concave function, regardless of whether the considered system is additive or not. The case of macrostates for which $S(U)$ is a convex function is much more complicated. In general, convex macrostates have less entropy than states represented by the concave envelope of $S(U)$. Therefore, such envelope states are realized in additive systems, where they correspond to phase separation. They are, however, forbidden in nonadditive systems. Thus, the lack of additivity forces the system to realize convex states, which (surprisingly) make the microcanonical ensemble much more interesting than the canonical one. An immediate consequence of such states is, for example, negative specific heat or negative magnetic susceptibility. Other non-common behaviours arising from ensemble nonequivalence relate to the entire spectrum of microcanonical phase transitions, which are not visible in other ensembles. 

In particular, a number of generic situations, regarding concave and convex entropies, and having the hallmarks of microcanonical phase transitions was discussed in Ref.~\cite{2005JStatPhysBouchet}. Many of these possible generic situations have not yet been observed in any models, not to mention real systems. Clearly, there is a lot of work to be done in this filed, especially since so far little attention has been paid to the partial equivalence of ensembles \cite{2007PhysACasetti}. To be honest, the preliminary classification of phase transitions in nonadditive \cite{2002PhysATouchette} systems developed in Ref.~\cite{2005JStatPhysBouchet} completely ignores the cases with locally affine entropy and focuses only on singularities arising in its concave and convex regions. This paper aims to make a small contribution to this omitted area.

In what follows, we study the partial equivalence of ensembles in the so-called \textit{minimal, diffusion-based spin model}, which has been introduced in Ref.~\cite{2016aPREFronczak}. It was already shown that the model exhibits very interesting critical-like behaviour (i.e. Thouless effect \cite{2016aPREFronczak} and hybrid phase transition \cite{2016bPREFronczak}), when it is analysed in the canonical and grand canonical ensemble. In this paper, we confront microcanonical and canonical properties of the model. Our study reveals that in this simple model a unique first order transition in both ensembles emerges, which results from affine thermodynamic potentials. Not so long ago, the possibility of such a behaviour was predicted theoretically \cite{2006PhysATouchette}. Such a generic situation, however, has not been raised in the context of phase transitions in partially equivalent ensembles \cite{2005JStatPhysBouchet}, which we are doing here.

\section{Partial equivalence of ensembles: case study}

\subsection{The model}

The spin model we deal with is completely defined by the Hamiltonian: 
\begin{equation}\label{H}
\mathcal{H}_{\!{_N}}(\Omega)=-N_{+}(\Omega)\ln a+\ln{N\choose N_{+}(\Omega)},
\end{equation}
where $\Omega=(s_1,s_2,\dots,s_N)$ represents microscopic configuration of the system of $N$ distinguishable spins $s_i=\pm 1$, with $N_+(\Omega)$ standing for the number of positive spins, and $a>0$ being the model external parameter (with notion from \cite{2016aPREFronczak} we take $a=(1-q)/q$). Although in Ref.~\cite{2016aPREFronczak}, the model was designed to provide theoretical explanation for certain critical-like phenomena observed in a dynamic, social network \cite{EPLLiu2012,PREBassler2015}, and although it was shown there that its dynamical properties can be one-to-one related to the phenomenon of the biased random walk, in this paper, we cut ourselves off from the question of whether the model is physically realistic or not. We just treat it as a toy model having some non-trivial properties resulting from the lack of additivity (i.e. $\mathcal{H}_{_N}+\mathcal{H}_{_R}\neq\mathcal{H}_{_{N\!+\!R}}$).

\subsection{Microcanonical ensemble}\label{SecMicro}

\begin{figure*}[t]
	\includegraphics[width=2\columnwidth]{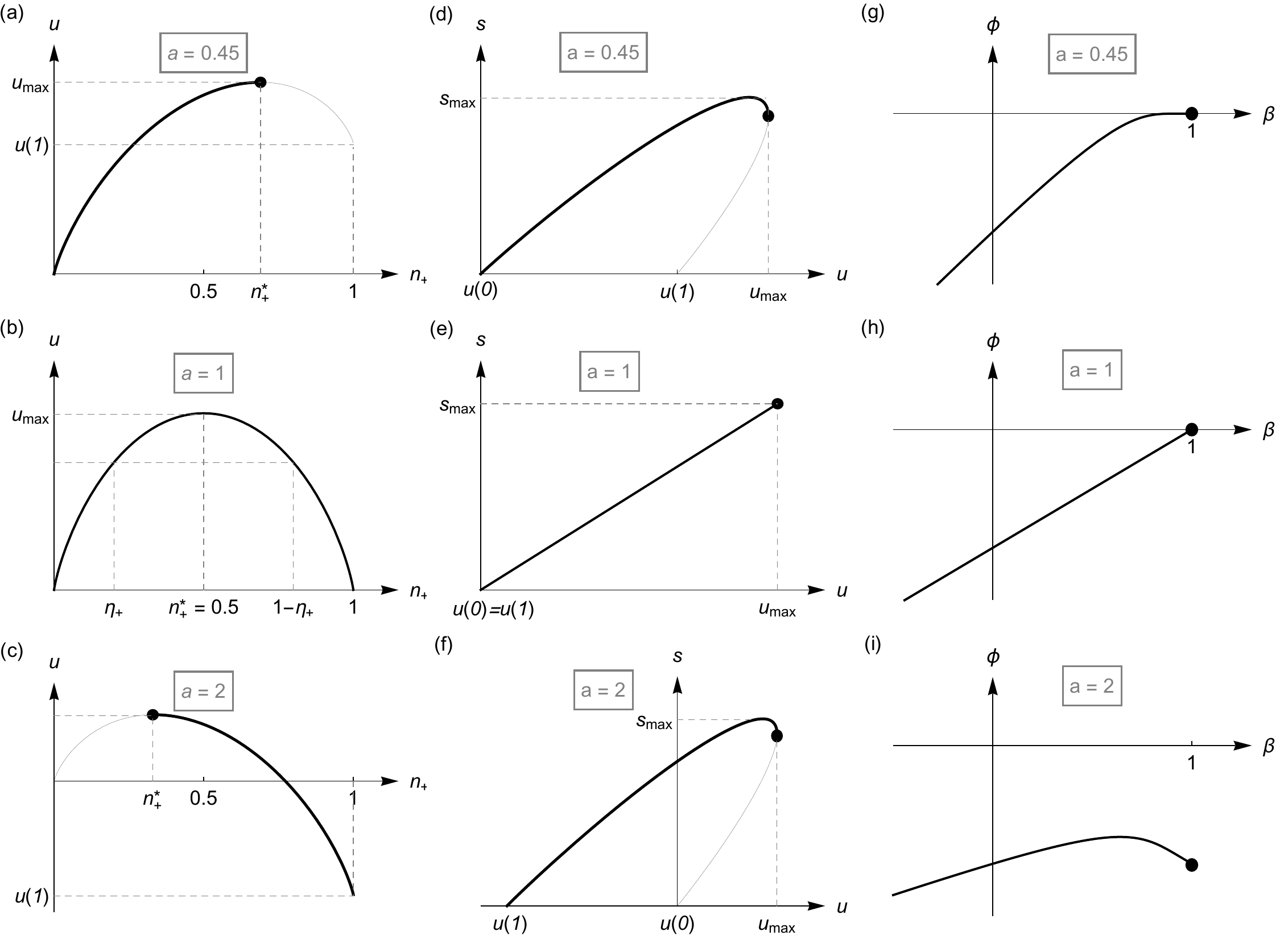}
	\caption{\label{fig1} \textbf{A few realizations of different thermodynamic state functions characterizing the model} for three representative values of the parameter $a=0.45,\;1.0,\;2.0$. In the first column, in graphs (a,b,c), internal energy, $u$, as a function of the number of positive spins, $n_{_+}$, is shown (cf.~Eq.~(\ref{unp})). In the second column, in graphs (d,e,f), microcanonical entropy per spin, $s$, is presented as a function of the internal energy, $u$ (see~Eq.~(\ref{su})). In the third column, in graphs (g,h,i), free energy per spin $\phi$ characterizing the model in the canonical ensemble is drawn as a function of $\beta$ (cf.~Eq.~(\ref{phi2})). The meaning of bold parts of the $u(n_{_+},a)$ and $s(u,a)$ curves, and other symbols in this figure (e.g. $u_{max}$, $n^*_{_+}$, etc.) is explained in the main text.}
\end{figure*}

\begin{figure*}[t]
	\includegraphics[width=1.8\columnwidth]{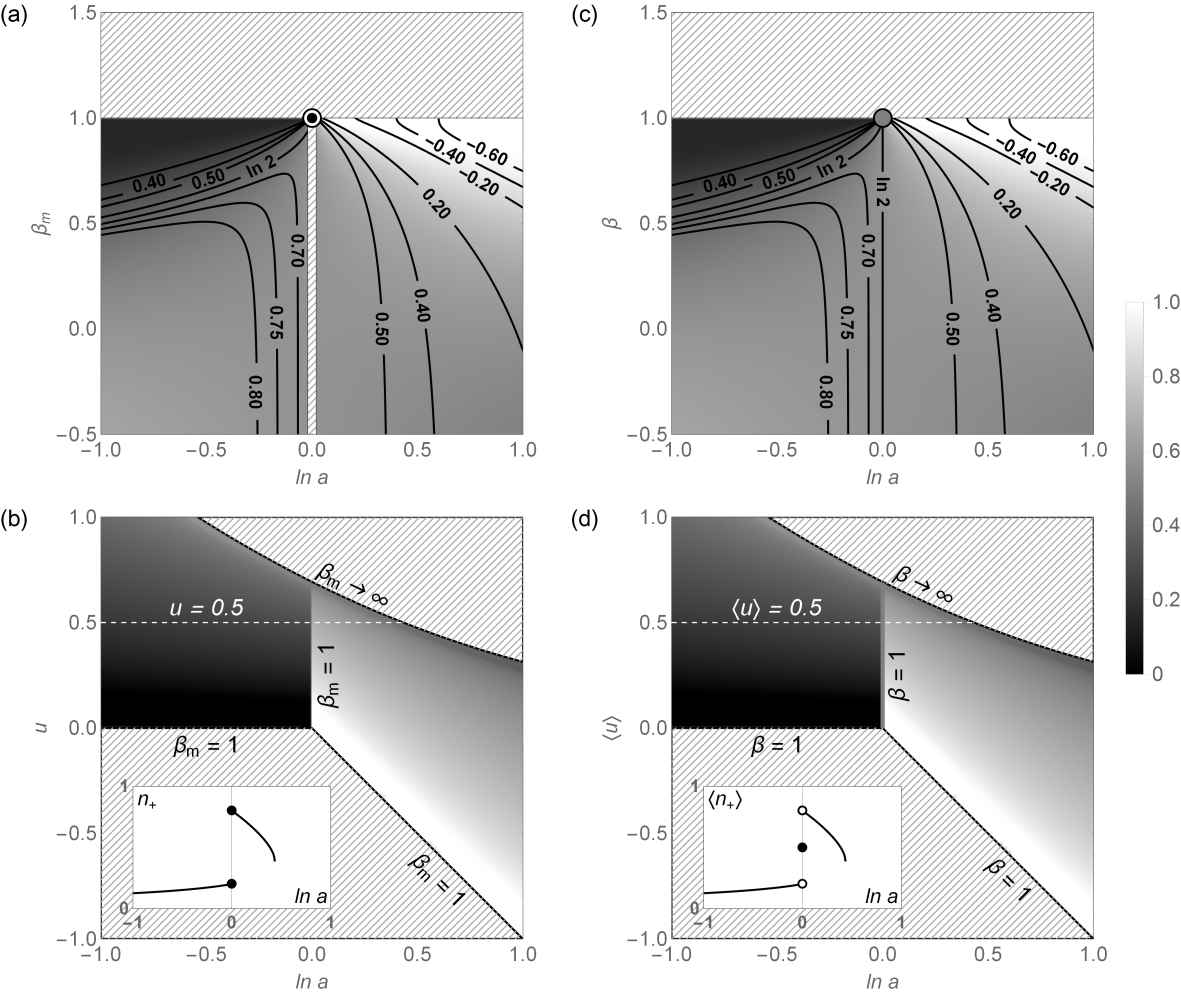}
	\caption{\label{fig2} \textbf{Phase diagrams of the model} in the microcanonical (a,b) and canonical ensemble (c,d). Two variants of phase diagrams are shown: i. (a) and (c) present the number of positive spins ($n_{_+}$ and $\langle n_{_+}\rangle$, cf.~Eqs.~(\ref{npm0}) and~(\ref{npc})) as a function of the inverse temperature ($\beta_{_m}$ and $\beta$, respectively) and the parameter $a$, and ii. (b) and (d) show the number of positive spins as a function of the internal energy (correspondingly, $u$ and $\langle u\rangle$) and the parameter $a$. The grayscale bar on the right represents the average number of spins. The hatchet areas indicate forbidden macrostates. The black contour lines in graphs (a) and (c) highlight states with a given value of the internal energy, $u$ and $\langle u\rangle$, respectively. The phase transition points: $a=1$ and $\beta_{_m}=\beta=1$, in diagrams (a) and (c), correspond to straight line segments: $a=1$ and $u,\langle u\rangle\in[0,\ln 2]$ in diagrams (b) and (d). Inset plots in (b) and (d) show cross-sections of these diagrams for $u=\langle u\rangle=0.5$, respectively.}
\end{figure*}

In textbooks on statistical physics, the discussion of the microcanonical ensemble usually precedes the discussion of the canonical ensemble. In this article, we also uphold this habit. Our first goal is to draw up the microcanonical phase diagram of the considered model and to identify phase transition points on it.

In the microcanonical ensemble, one assumes that the energy $U$ of the system as a whole is fixed. In this paper, however, since we study systems in the thermodynamic limit, $N\rightarrow\infty$, instead of the total energy $U$, as the control parameter, we use the energy per spin, $u$, which is given by, cf.~Eq.~(\ref{H}),
\begin{equation}\label{unp}
u=\frac{U}{N}=-\ln a\:n_{_+}-n_{_+}\ln n_{_+}-n_{_-}\ln n_{_-},
\end{equation}
where
\begin{equation}\label{nppom}
n_{_+}=N_{_+}/N,\;\;\;\;\;\mbox{and}\;\;\;\;\;n_{_-}=1-n_{_+}.
\end{equation}
Correspondingly, we also use the notion of the entropy per spin
\begin{eqnarray}\label{su}
s=\frac{S}{N}=\frac{\ln {N\choose N_{_+}}}{N}&=&-n_{_+}\ln n_{_+}-n_{_-}\ln n_{_-}\\\label{su1}&=&u+n_{_+}\ln a.
\end{eqnarray}

Eq.~(\ref{unp}) defines energy as a function of the number of positive spins $n_{_+}$ and the parameter $a$, i.e. $u(n_{_+},a)$. Since the variable $n_{_+}$ belongs to the range $[0,1]$, the energy per spin is not arbitrary, but it meets certain restrictions (see Fig.~\ref{fig1}~(a,b,c)). In particular,
\begin{itemize}
\item[for] $\;a<1$: $\;\;u\in[u(0),u_{max}]\!=\![0,\ln\frac{1+a}{a}]$,
\item[for] $\;a=1$: $\;\;u\in[u(0),u_{max}]\!=\![u(1),u_{max}]\!=\![0,\ln 2]$,
\item[for] $\;a>1$: $\;\;u\in[u(1),u_{max}]=[-\ln a,\ln\frac{1+a}{a}]$,
\end{itemize}
where $u_{max}=u(n_{_+}^*,a)=\ln\frac{a+1}{a}$ is the maximum energy value obtained for $n_{_+}^*=\frac{1}{a+1}$. A few realizations of $u(n_{_+},a)$, for various representative values of the parameter $a$, are shown in Fig.~\ref{fig1}~(a,b,c). In this figure, one can see that certain values of $u$ can be realized in two ways, i.e. for two different values of~$n_{_+}$. The bold parts of the $u(n_{_+},a)$ curves represent those system realizations (macrostates) that are more likely to occur and therefore they have greater entropy. For clarity, in~Fig.~\ref{fig1} (d,e,f), such maximum-entropy states are also marked as bold parts of the $s(u,a)$ curves. By making simple calculations \footnote{The function $f(x)$ is strictly concave when $\frac{\partial^2f}{\partial x^2}<0$. It is strictly convex, when $\frac{\partial^2f}{\partial x^2}>0$. Finally, $f(x)$ is affine, when $\frac{\partial^2f}{\partial x^2}=0$.}, one can show that these bold entropy branches are strictly concave for all values of the parameter~$a\neq 1$. The case, when $a=1$ is an exception. Then, entropy becomes affine in the whole energy range: $s(u,a)=u$ for $u\in[0,\ln 2]$. According to what was said at the beginning of this paper, the strict concavity of $s(u,a)$ proves the equivalence of statistical ensembles while affinity results in partial equivalence. Now, we will take a closer look at these issues. To this aim, we will compare phase diagrams: microcanonical and canonical, of the model. 

To start with, let us note that the microcanonical temperature of the model is given by:
\begin{equation}\label{betam}
\beta_{_m}=\frac{\partial s}{\partial u}=\frac{\partial s}{\partial n_{_+}} \frac{\partial n_{_+}}{\partial u}=\frac{\ln\left(\frac{n_{_+}}{n_{_-}}\right)} {-\ln a+\ln\left(\frac{n_{_+}}{n_{_-}}\right)},
\end{equation}
where $n_{_+}$ depends on $u$ and $a$, which are the control parameters in this ensemble. In particular, for $a=1$, regardless of the energy value, the microcanonical temperature is always $\beta_{_m}=1$. It means that phase diagram points for which $a\!=\!1$ and $\beta_{_m}\neq 1$ are forbidden, see Fig.~\ref{fig2}~(a). Correspondingly, for $a=1$ and $\beta_{_m}=1$, discontinuous phase transition occurs for different values of $u$. It is due to the fact that for $a=1$ (or $\ln a=0$) and for the energy values $u\in [0,\ln 2)$, there are two different macrostates (i.e. $\eta_{_+}$ and $1-\eta_{_+}$, Fig.~\ref{fig1}~(b)) having the same entropy. When $a\neq 1$, one of these states becomes more entropic. This leads to a discontinuous jump in $n_{_+}$. The size of this jump depends on the energy of the system. This is clearly visible in Fig.~\ref{fig2}~(b), where $n_{_+}$ is depicted as a function of $u$ and $a$.

Finally, from Eq.~(\ref{betam}), the expression for $n_{_+}$ as a~function of $\beta_{_m}$ and $a\neq 1$ can be obtained: 
\begin{eqnarray}\label{npm0}
n_{_+}(\beta_{_m},a)=\frac{1}{1+a^{\kappa_{_m}}},\;\;\;\;\mbox{where}\;\;\;\; \kappa_{_m}=\frac{\beta_{_m}}{\beta_{_m}-1}, 
\end{eqnarray}  
which is useful in drawing up the microcanonical phase diagram of the model. In the diagram, Fig.~\ref{fig2}~(a), forbidden macrostates are marked as hatched areas. In addition to states for which $a=1$ and $\beta_{_m}\neq 1$, all macrostates with $\beta_{_m}>1$ are also unavailable. The solid lines in Fig.~\ref{fig2}~(a) indicate states with the same energy (numbers placed on these curves correspond to different values of~$u$). The point $a=\beta_{_m}=1$, at which all these curves converge, is the point of the first order microcanonical transition. From Eqs.~(\ref{unp}) and~(\ref{npm0}) one can show that for 
\begin{eqnarray}\label{npm1}
\lim_{\beta_{_m}\rightarrow 1^{-}} n_{_+}(\beta_{_m},a)&=&\left\{ \begin{array}{ccl}
0 & \mbox{for } & \ln a<0 \\
\eta_{+} & \mbox{for } & \ln a\rightarrow 0^{-} \\
1-\eta_{+} & \mbox{for } & \ln a\rightarrow 0^{+} \\
1 & \mbox{for } & \ln a>0
\end{array}\right.,
\end{eqnarray}
where $\eta_{_+}\leq \frac{1}{2}$ represents the average number of spins in the system for which $a=1$. Of course, $\eta_{_+}$ depends on $u$, see Fig.~\ref{fig1}~(b) and Fig.~\ref{fig2}~(b).

Below we show that the canonical phase diagram differs from the just described microcanonical one.

\subsection{Canonical ensemble}\label{SecCan}

As mentioned at the beginning of this paper, some canonical analysis of the considered model has already been  carried out. Namely, in Ref.~\cite{2016aPREFronczak}, properties of the model at fixed temperature, $\beta=1$, have been studied. It was shown there that, when the parameter $a$ approaches unity, an interesting discontinuous phase transition with diverging response function is observed. 

To be concrete, for $\beta=1$, the partition function was shown to be given by:
\begin{equation}\label{pom1}
Z_N(a)=\sum_{\Omega}e^{-\mathcal{H}_{\!{_N}}(\Omega)}=\frac{1-a^{N+1}}{1-a}.
\end{equation} 
Resulting from the above expression, Helmholtz free energy per spin has a singularity at $a=1$. This singularity leads to a discontinuous jump in the average number of positive spins:
\begin{eqnarray}\label{pom2}
\langle n_{_+}\rangle&=&\lim_{N\rightarrow\infty}\frac{\langle N_{_+}\rangle}{N} =\left\{ \begin{array}{lcl}
0 & \mbox{for } & \ln a<0 \\
\frac{1}{2} & \mbox{for } & \ln a=0\\
1 & \mbox{for } & \ln a>0
\end{array}\right.,
\end{eqnarray}
which is accompanied by diverging susceptibility
\begin{eqnarray}\label{pom3}
\chi=\frac{\partial\langle n_{_+}\rangle}{\partial a}\sim \frac{1}{N}|a-1|^{-2}.
\end{eqnarray}
Although further in this section, a more extensive canonical analysis of this model for any temperature value is performed, already at this moment, we highlight that at the transition point: $\beta=\beta_{_m}=1$ and $a=1$, both (microcanonical and canonical) ensembles differ from each other at the level of macrostates, cf.~Eqs.~(\ref{npm1}) and (\ref{pom2}).

For arbitrary temperature, in the continuum limit, the partition function of the model can be written as:
\begin{eqnarray}\label{Z1a}
Z_N(\beta,a)&=&\sum_{\Omega}e^{-\beta\mathcal{H}_{\!{_N}}(\Omega)}\\\label{Z1b}
&=&\!\sum_{N_{_+}=0}^N {N\choose N_{_+}}^{1-\beta}a^{\beta N_{_+}}\\\label{Z1c}
&=&\!\int_{0}^1 e^{-N \phi(n_+;\beta,a)}dn_{_+}\!,
\end{eqnarray} 
where 
\begin{eqnarray}\nonumber
\phi(n_{_+};\beta,a)&=&\left(1-\beta)(n_{_+}\ln n_{_+}+ 
n_{_-}\ln n_{_-}\right)\\\label{phi1}&&-\beta\:n_{_+}\ln a.
\end{eqnarray} 
It is easy to check that for $\beta<1$ the function $\phi(n_{_+};\beta,a)$ is strictly convex with respect to $n_{_+}$ on the interval $[0,1]$. Furthermore, in this interval, it has a minimum at,
\begin{eqnarray}\label{npc}
n_{_+}^c=\frac{1}{1+a^\kappa},\;\;\;\;\mbox{where}\;\;\;\;
\kappa=\frac{\beta}{\beta-1}.
\end{eqnarray}  
Therefore, for $\beta<1$, the Laplace method of steepest descents (a.k.a. the saddle-point method) can be used to calculate the integral (\ref{Z1c}). In this way one gets: 
\begin{eqnarray}\label{Z2}
Z_N(\beta,a)&\asymp& e^{-N \phi(\beta,a)},
\end{eqnarray} 
where
\begin{eqnarray}\label{phi2}
\phi(\beta,a)\stackrel{Eq.(\ref{phi1})}{=} \phi(n_{_+}^c;\beta,a)=(1\!-\!\beta) \ln\left(\frac{a^\kappa}{1+a^\kappa}\right).
\end{eqnarray} 
The symbol '$\asymp$' in Eq.~(\ref{Z2}) means that, as $N\rightarrow\infty$, the dominant part of the partition function, $Z_N(\beta,a)$, scales exponentially with the system size. 

Rephrasing the remark closing the last paragraph, one could say that for $\beta<1$ the considered system is extensive. This is because its free energy \footnote{It should be clarified that, in the canonical ensemble, the Helmholtz free energy is defined as $F_{\!_N}(\beta,a)=\Phi_{\!_N}(\beta,a)/\beta$. However, compared with $F_{\!_N}(\beta,a)$, some features of $\Phi_{\!_N}(\beta,a)$ make the latter function a better choice, especially when it comes to studies on the equivalence of statistical ensembles \cite{2005JStatPhysBouchet, 2009PhysRepTouchette}.}
\begin{equation}\label{Phi1}
\Phi_{\!_N}(\beta,a)=-\ln Z_N(\beta,a)\stackrel{N\gg 1}{=}N\phi(\beta,a),
\end{equation}
is linear with respect to $N$. This confirms the known fact that extensiveness is not in contradiction with the lack of additivity. On the other hand, for $\beta>1$, due to strict concavity of $\phi(n_{_+};\beta,a)$ with respect to $n_{_+}$, the model is non-extensive. We will not bother with this case here.

In what follows, we discuss thermodynamic properties of the model in the canonical ensemble. These properties can be easily to obtained from the bulk free energy, $\phi(\beta,a)$~(\ref{phi2}), which is an analytic and concave function for all: $\beta<1$ and $a>0$, see~Fig.~\ref{fig1}~(g,h,i). For these parameter ranges, the average number of positive spins is also an analytic function,
\begin{eqnarray}\label{nps}
\langle n_{_+}\rangle\!=\frac{1}{\beta}\frac{\partial\phi(\beta,a)} {\partial\ln a}\!=\frac{1}{1+a^\kappa},
\end{eqnarray}
just like the average energy
\begin{eqnarray}\label{us}
\langle u\rangle\!=\frac{\partial\phi(\beta,a)} {\partial\beta}\!=\!
-\langle n_{_+}\rangle\ln\left(a\langle n_{_+}\rangle\right)-\langle n_{_-}\rangle \ln\langle n_{_-}\rangle.
\end{eqnarray}
where $\langle n_{_-}\rangle=1-\langle n_{_+}\rangle$. It is remarkable, that Eqs.~(\ref{nps}) and~(\ref{us}) have the same form as the corresponding Eqs.~(\ref{npm0}) and~(\ref{unp}) in the microcanonical ensemble. The difference, however, arises when ranges of their applicability are taken into account. Namely, in the canonical ensemble, when $a=1$, for the entire temperature range, $\beta\leq 1$, one has: $\langle n_{_+}\rangle=\frac{1}{2}$ and $\langle u\rangle=\ln 2$. On the other hand, in the microcanonical ensemble, when $a=1$, the states with $\beta_{_m}<1$ are forbidden, whereas for $\beta_{_m}=1$ the number of positive spins, $\eta_{_+}$, depends on $u\in[0,\ln 2]$, see Eq.~(\ref{npm1}). The above properties of the model are illustrated in Fig.~\ref{fig2}, where the canonical phase diagram is shown in comparison to its microcanonical version. 

\begin{figure}[t]
	\includegraphics[width=0.9\columnwidth]{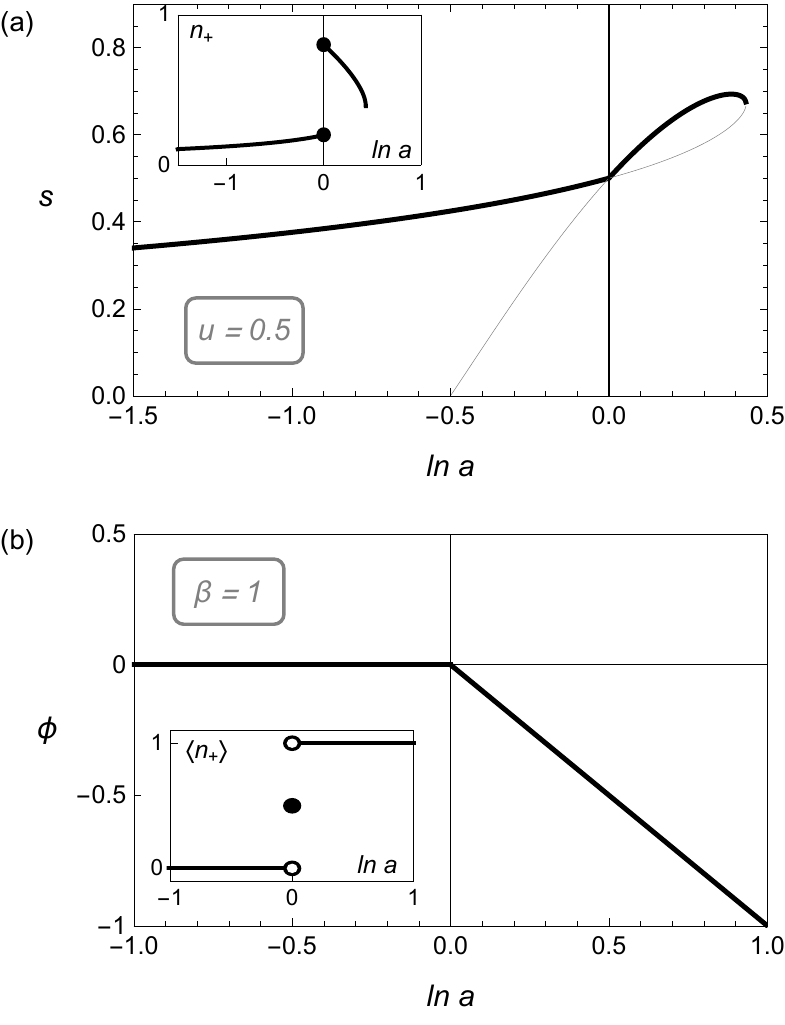}
	\caption{\label{fig3} \textbf{Thermodynamic state functions in the vicinity of the transition point}: (a) the microcanonical energy per spin, $s(u,a)$ (\ref{su1}), as a function of the parameter $a$ for $u=0.5$, and (b) the free energy per spin, $\phi(\beta,a)$ (\ref{phi2}), as a function of the parameter $a$ for $\beta=1$. These state functions have non-analytical points for $a=1$ and $u\in[0,\ln 2)$, in the microcanonical ensemble, and for $a=\beta=1$, in the canonical ensemble, respectively. These non-analytical points are seen as discontinuities of  the corresponding derivatives: $n_{_+}=\partial s/\partial\ln a$ and $\langle n_{_+}\rangle=-\partial\phi/\partial\ln a$ (see inset plots).}
\end{figure}

\subsection{Partial equivalence and phase transitions}
 
The equivalence of microcanonical and canonical ensembles of the model studied, which holds for $\beta\leq 1$ and $a\neq 1$, is in agreement with theoretical predictions of Ref.~\cite{2000JStatPhysEllis}. It is due to strict concavity of the microcanonical entropy as a function of $u$. For $a=1$, when entropy becomes affine in $u$, partial equivalence is observed in the model, again in agreement with Ref.~\cite{2000JStatPhysEllis}. 

As emphasized in Ref.~\cite{2007PhysACasetti}, in a situation of partial equivalence, a whole set of values of the control parameter in one ensemble corresponds to a single value of the control parameter in the other ensemble. In particular, the same value of the average energy can be obtained for the whole set of canonical temperatures, or the whole set of energies characterizing  microcanonical systems may show the same microcanonical temperature. In our case study, both behaviors are observed, which make the considered model interesting, because in the systems studied so far, at most one (usually the first one) of these behaviors was observed.

Another valuable feature of the model studied is a unique phenomenology of the observed first order transition that occurs simultaneously in both ensembles. In additive systems, the commonly known cause of such simultaneous transitions is the bimodal shape of the microcanonical entropy as a function of the energy. In nonadditive systems, when only strict nonequivalence of ensembles is taken into account, much richer phenomenology comes into play \cite{2005JStatPhysBouchet}, which, however, in any of possible scenarios, does not lead to simultaneous (in both ensembles) first order transitions. Our study shows that such a scenario is possible in partially equivalent ensembles. 

In our model, the microcanonical transition arises as a result of singularity in entropy $s(u,a)$ (\ref{su}) that appears at $a=1$ (and correspondingly $\beta_{_m}=1$) when energy of the system is in the range $u\in[0,\ln 2)$. At this point, the right- and left-sided derivatives of $s(u,a)$ with respect to $\ln a$, which give the average number of positive spins, $n_{_+}=\partial s/\partial\ln a$, differ from each other, see Eq.~(\ref{npm1}) and~Fig.~\ref{fig3}(a). On the other hand, the canonical first order transition, that appears for the same parameter values as in the microcanonical ensemble, $a=1$ and $\beta=1$, results from the free energy~(\ref{phi2}) composing of two parts: flat (i.e. $\phi(1,a)\stackrel{a<1}{=}0$) and affine (i.e. $\phi(1,a)\stackrel{a\geq 1}{=}-\ln a$), at the interface of which a nonanalytical point develops, see Eq.~(\ref{pom2}) and Fig.~\ref{fig3}(b). Not so long ago, purely theoretical considerations about such a canonical transition have been putted forward in Ref.~\cite{2006PhysATouchette}. To our knowledge, the model studied in this paper is the first one in which such a transition was observed as resulting from the affine microcanonical entropy which is defined on a limited energy range.

\section{Concluding remarks}

In this paper, we draw attention to the problem of phase transitions in partially equivalent ensembles. We focus on a very simple spin model which shows interesting discontinuous phase transitions that are simultaneously observed in all three basic statistical ensembles (microcanonical, canonical, and grand canonical), although in each of these three ensembles they occur in a slightly different way.

In the microcanonical ensemble (see Sec.~\ref{SecMicro}), the transition is associated with a discontinuous change in the average number of positive spins (which is the order parameter of the transition), that is not accompanied by any fluctuations. In the canonical ensemble (see Sec.~\ref{SecCan} and Ref.~\cite{2016aPREFronczak}), the average number of spins also changes discontinuously, but this change is accompanied by power-law fluctuations, which are, however, size-dependent and disappear in the thermodynamic limit. Finally, in the grand canonical ensemble (see Ref.~\cite{2016bPREFronczak}), the phase transition is mixed-order (hybrid), because the discontinuous change of the order parameter is accompanied by diverging (i.e. critical-like) fluctuations.

The unique nature of phase transitions observed in this model causes that it escapes the existing classifications of these phenomena. (Here, we mean not only the well-established 'modern' classification of phase transitions, which applies to additive systems, but also the recent attempt to generalize this classification into non-additive systems that was introduced in \cite{2005JStatPhysBouchet}.) In a broader context, results reported in this paper show that existing classifications must be refined. Particularly, the issue of phase transitions in partially equivalent ensembles (with locally affine microcanonical entropy) needs deeper insight. It also needs to be clarified, whether the coexistence of affine microcanonical entropy and hybrid transition (see also \cite{2014PRLBar,2014JStatMechBar}) that is observed in this model is accidental or not. 


\section{Acknowledgments}

This work has been supported by the National Science Centre of Poland (Narodowe Centrum Nauki, NCN) under grant no. 2015/18/E/ST2/00560 (A.F. and G. S.).

\end{document}